\documentclass[a4paper]{article}

\usepackage{17lomcon}        
\usepackage{cite}             
\usepackage{epsfig}           

\begin{document}


\title{THE PLANCK LEGACY \\ REINFORCING THE CASE FOR A STANDARD MODEL OF COSMOLOGY: $\Lambda$CDM}

\author{Nazzareno Mandolesi$^{1,5,2}$ \email{mandolesi@iasfbo.inaf.it}, Diego Molinari$^{2,1}$ \email{molinari@iasfbo.inaf.it}, Alessandro Gruppuso $^{1,3}$ \email{gruppuso@iasfbo.inaf.it}, Carlo Burigana$^{1,2,3}$ \email{burigana@iasfbo.inaf.it} and Paolo Natoli$^{2,4,1}$ \email{paolo.natoli@gmail.com} \\ (on behalf of {\it Planck} Collaboration)}

\affiliation{$^{1}$INAF--IASF \footnote{Istituto Nazionale di Astrofisica -- Istituto di Astrofisica Spaziale e Fisica Cosmica}Bologna, Via Piero Gobetti 101, I-40129 Bologna, Italy \\
$^{2}$Dipartimento di Fisica e Scienze della Terra, Universit\`a degli Studi di Ferrara,\\
Via Giuseppe Saragat 1, I-44122 Ferrara, Italy\\
$^{3}$INFN \footnote{Istituto Nazionale di Fisica Nucleare}, Sezione di Bologna, Via Irnerio 46, I-40126, Bologna, Italy\\
$^{4}$Agenzia Spaziale Italiana Science Data Center, \\ Via del Politecnico snc, 00133, Roma, Italy\\
$^{5}$Agenzia Spaziale Italiana, Viale Liegi 26, Roma, Italy}


\date{}
\maketitle


\begin{abstract}
   We present a brief review of the main results of the {\it Planck} 2015 release describing the new calibration of the data, showing the maps delivered in temperature and, for the first time, in polarization, the cosmological parameters and the lensing potential. In addition we present a forecast of the Galactic foregrounds in polarization. Future satellite experiments will have the challenge to remove the foregrounds with great accuracy to be able to measure a tensor-to-scalar ratio of less than 0.01.
\end{abstract}

\section{Introduction}

Cosmic Microwave Background (CMB) radiation, whose first prediction can be attributed to Ralph Alpher, Robert Herman and George Gamov in 1948 \cite{Alpher1948}, is a blackbody radiation with T=$2.72548 \pm 0.00057$ K\cite{Fixsen2009} extremely uniform across the whole sky. It is the relic radiation emitted at the time the nuclei and electrons recombined to form neutral hydrogen, when the Universe was about 400,000 years old and the mean free path of the photons became larger than the Universe itself. The CMB hypothetical emission surface is known as Last Scattering Surface (LSS). The CMB tiny temperature and polarization anisotropies encode a wealth of cosmological information.

The {\it Planck} mission, launched and operated by ESA, represent a third-generation satellite devoted to the measurement of the temperature and polarization anisotropies of the CMB with unprecedented precision, to test the standard cosmological model and investigate the early universe physics. {\it Planck} has observed the microwave sky continuously from the 12th of August 2009 to the 23rd of October 2013 in 9 bands between 30 GHz and 1 THz with angular resolutions between 5 and 30 arcmin and a sensitivity of $\Delta T/T_{CMB} \sim 2 \times 10^{-6}$ \cite{Tauber2010}. 
The satellite is composed by two instruments: a Low Frequency Instrument (LFI), with pseudo-correlation radiometers observing at 30, 44, 70 GHz that operated for about 48 months ,and an High Frequency Instrument (HFI), with bolometers observing at 100, 143, 217, 353, 545 and 857 GHz that operated for about 30 months. 
A first cosmological release has been delivered in 2013 \cite{PlanckI2013} in which the first 15 months of temperature data have been analysed. A second cosmological release occurred in 2015 \cite{PlanckI2015} involving the full mission data in temperature and, for the first time, including the polarization data.

In the present work we present a brief review of the main results of the {\it Planck} 2015 release.

\section{Calibration}

\begin{table}
\centering
\begin{tabular}{|cccc|}
\hline
 & & \multicolumn{2}{c|}{Galactic coordinates} \\
 & Amplitude & l & b \\
Experiment & $\mu \rm K_{\rm CMB}$ & [deg] & [deg] \\
\hline
LFI & $3365.5 \pm 3.0$ & $264.01 \pm 0.05$ & $48.26 \pm 0.02$ \\
HFI & $3364.5 \pm 1.0$ & $263.94 \pm 0.02$ & $48.21 \pm 0.008$ \\
{\it Planck} 2015 nominal & $3364.5 \pm 2.0$ & $264.00 \pm 0.03$ & $48.24 \pm 0.02$ \\
WMAP & $3355 \pm 8$ & $263.99 \pm 0.14$ & $48.26 \pm 0.03$ \\
\hline
\end{tabular}
\caption{LFI, HFI, and WMAP measurements of the Solar dipole\cite{PlanckI2015}.}
\label{tab:calibration}
\end{table}

In the {\it Planck} 2013 release, the photometric calibration of the data has been performed using the Solar dipole, that is the dipole induced in the CMB by the motion of the Solar System barycentre with respect to the LSS.
In the {\it Planck} 2015 release, instead, the photometric calibration has been performed using the orbital dipole. This is the modulation of the Solar dipole induced by the orbital motion of the satellite around the Solar System barycentre. Using this method, we can extract an independent measurement of the Solar dipole for each detector and use them in the {\it Planck} calibration pipeline. Since the orbital motion of the satellite is very well known, the measurement of the orbital dipole is one of the most accurate calibration methods. 

The amplitude of the orbital dipole is one order of magnitude smaller than the Solar dipole and therefore it is important to have a low noise and a good control of the systematic effects. A measurement of the Solar dipole gives an a posteriori check on calibration within the two {\it Planck} instruments. This is fundamental for a joint analysis of the data, and for the comparison with other CMB experiments. In Table \ref{tab:calibration} we report the amplitude and direction of the Solar dipole for the LFI \cite{PlanckV2015} and HFI \cite{PlanckVIII2015} instruments in comparison with WMAP \cite{Hinshaw2009}. The {\it Planck} nominal dipole is the result of the combination of the LFI and HFI measurements and it is used to carry out subtraction of the dipole from the frequency maps.
The measurements of the Solar dipole independently performed by {\it Planck} and WMAP agree within 0.28\% in amplitude, and to better than 2 arcmin in direction, proving the good consistency between the two experiments and their good calibration.

\section{Maps and power spectrum}

\begin{figure}[t]
     \centering
     \includegraphics[scale=0.12]{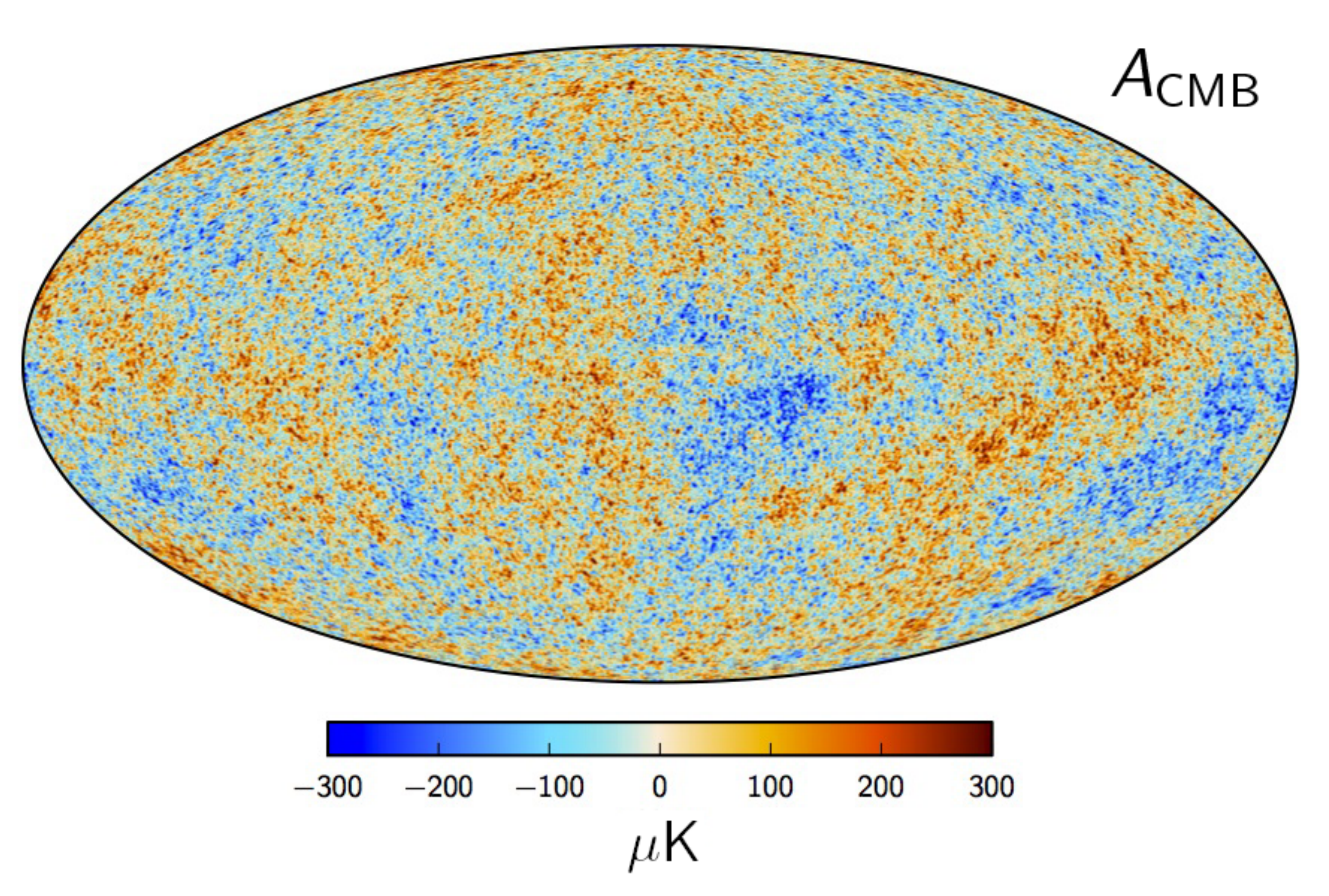}
     \includegraphics[scale=0.5]{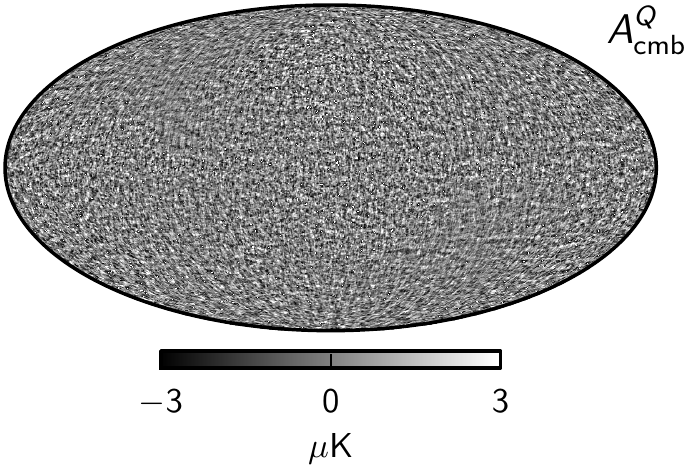}
     \includegraphics[scale=0.5]{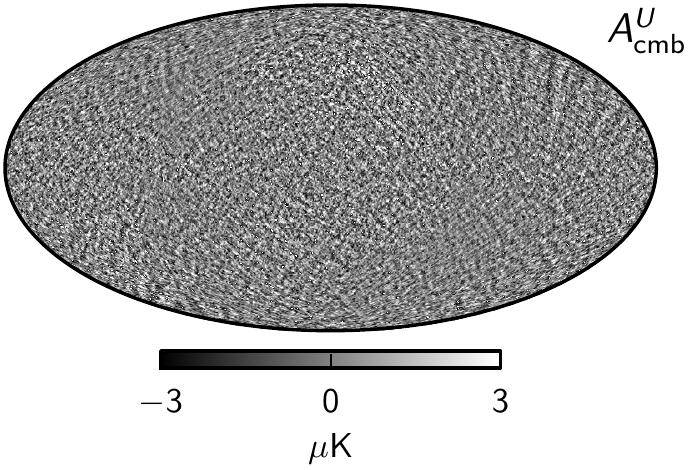}

\caption{({\it top}) CMB total intensity map at 5 arcmin from a joint analysis of the {\it Planck}, WMAP and 408 MHz observations. ({\it bottom}) Stokes Q ({\it left}) and U ({\it right}) amplitude maps results of the component separation of the {\it Planck} data\cite{PlanckIX2015,PlanckX2015}.}
\label{Maps}
\end{figure}

Four different component separation methods have been considered in the {\it Planck} collaboration to separate the contribution of the CMB signal from all the astrophysical contaminants emitted by our Galaxy or with an extragalactic origin \cite{PlanckIX2015}. The use of multiple methods is needed to have a robustness check of the products of the component separations. The methods are: SMICA \cite{Cardoso2008} (independent component analysis of power spectra), NILC \cite{Delabrouille2009} (needlet-based internal linear combination), Sevem \cite{Fernandez-Cobos2012} (template fitting) and Commander \cite{Eriksen2008} (pixel-based parameter and template fitting with Gibbs sampling). All the methods have been applied to the {\it Planck} data producing four sets of cleaned CMB maps in temperature and polarization. An example of the results is shown in Fig. \ref{Maps}. SMICA and Commander methods produce also maps of the foregrounds removed, useful for astrophysical studies. 

The temperature maps delivered by the four methods are very consistent with each other and are expected to give equivalently robust results when used for cosmological analyses, e.g. tests of isotropy, gaussianity and estimation of the lensing effect. The {\it Planck} 2015 release is characterised by the first release of polarization data and maps. The cleaned polarization CMB maps represent an important improvement in terms of coverage, angular resolution, and sensitivity with respect to previous results. However, the maps still show the presence of important anomalous features at large angular scales, due to systematic effects in the input frequency maps between 100 and 217 GHz. Since their analysis is still on-going, the first multipoles ($\ell < 30$) have been filtered out from the polarization maps delivered in 2015.

\begin{figure}[t]
     \includegraphics[scale=0.21]{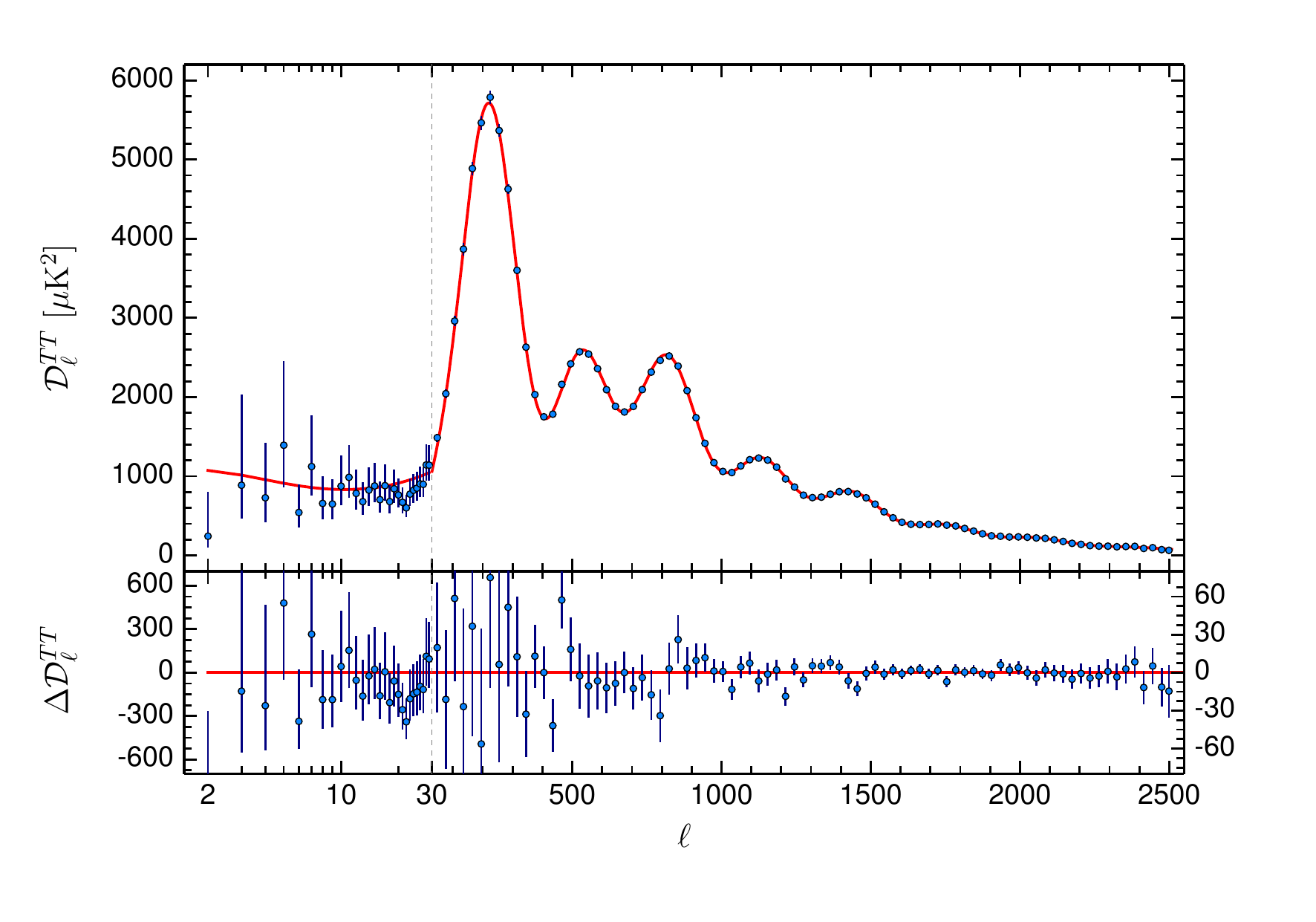}
     \includegraphics[scale=0.21]{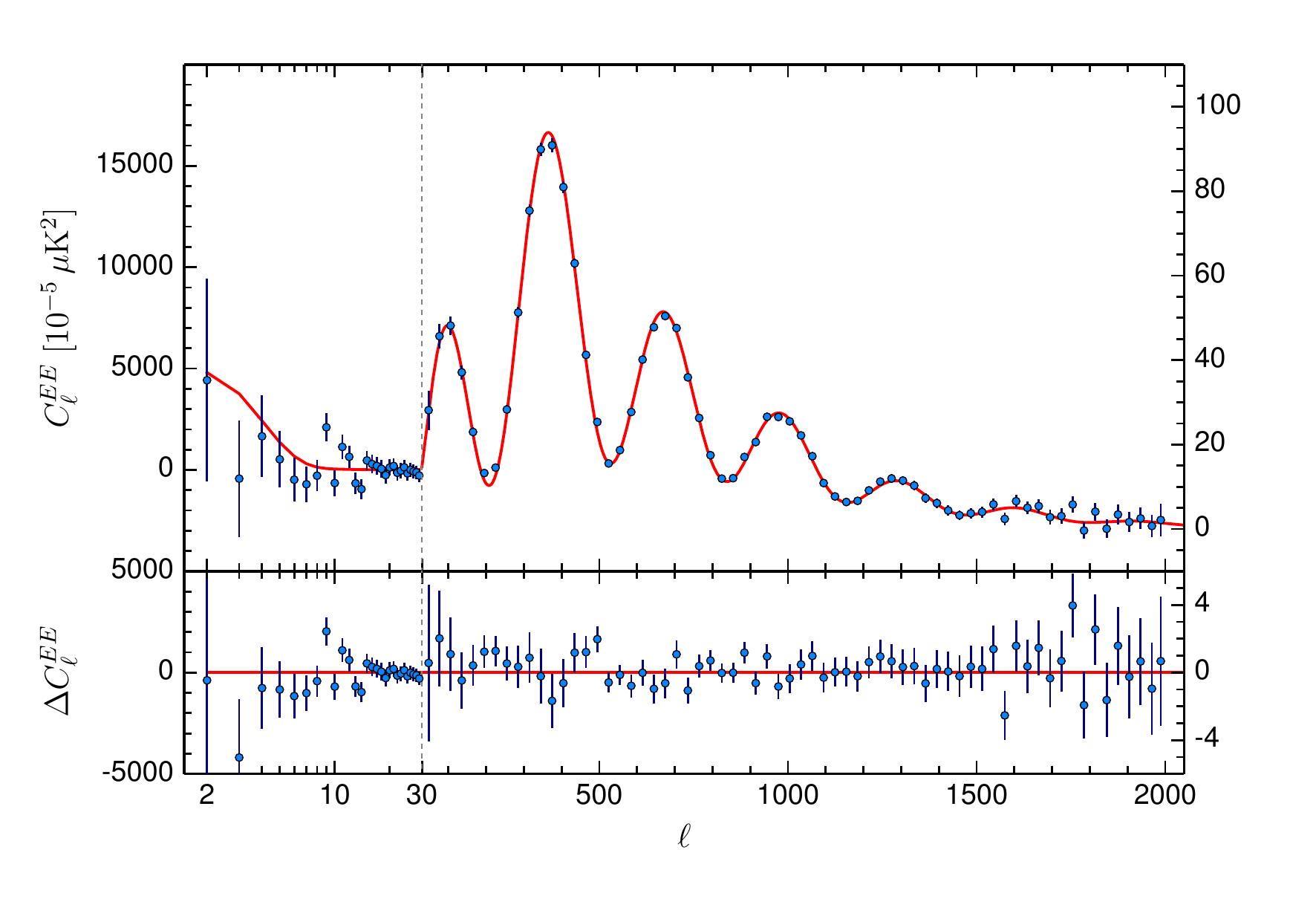}
     \includegraphics[scale=0.21]{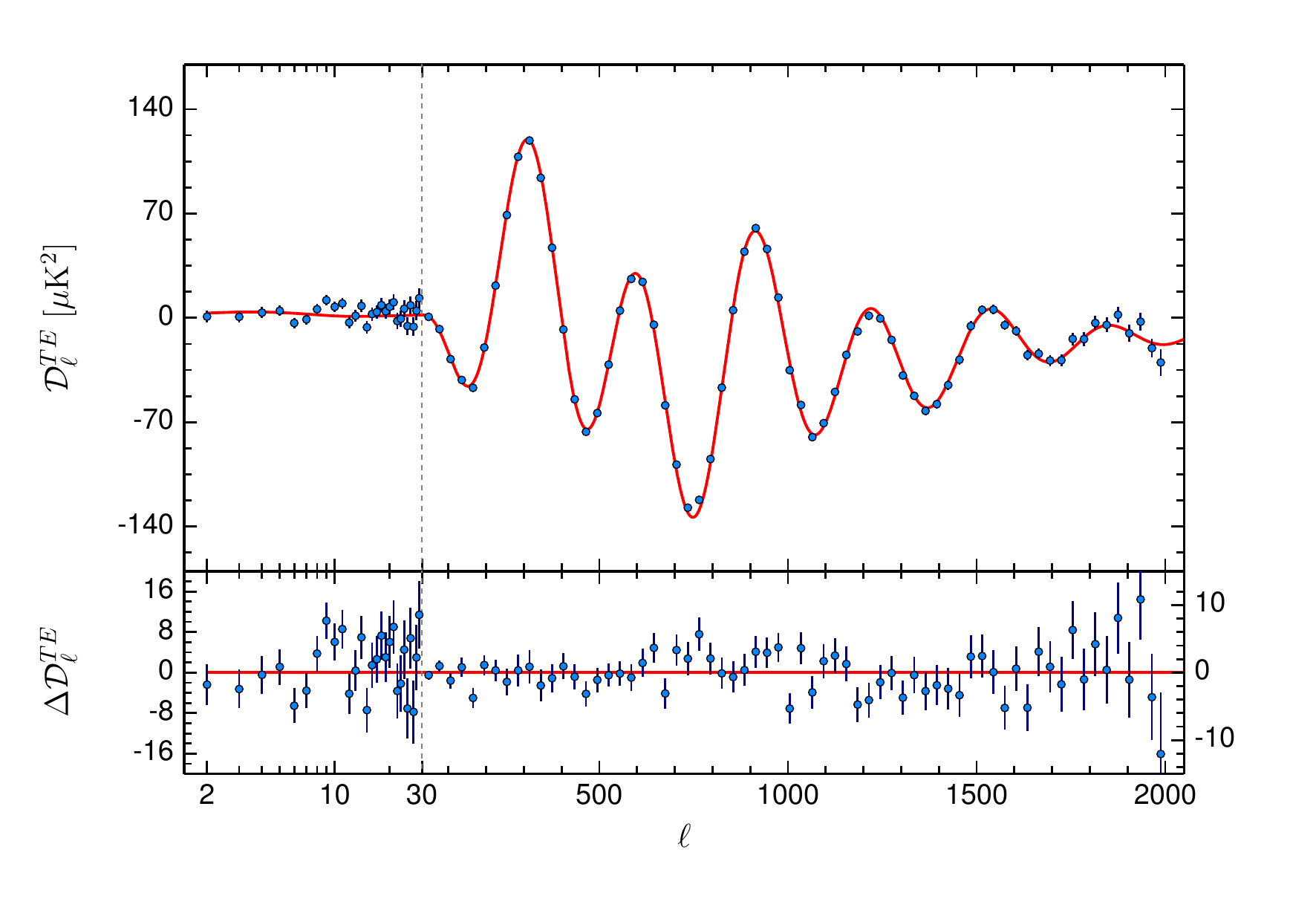}
\caption{The {\it Planck} 2015 angular power spectra in temperature ({\it left}) and polarization, EE ({\it center}) and TE ({\it right}) and their residuals with respect to the theoretical model shown in red obtained from only temperature and large scales ($\ell < 30$) polarization data\cite{PlanckXI2015}.}
\label{Spectra}
\end{figure}

In the assumption that the fluctuations of the anisotropy pattern are purely gaussian, all the information present in the maps can be encoded in the two point correlation function or in its harmonic transform, the Angular Power Spectrum (APS). Other {\it Planck} analyses \cite{PlanckXVI2015,PlanckXVII2015} did not find any significant deviation from gaussianity supporting this assumption. In Fig. \ref{Spectra} we show the APS in temperature and polarization and their residuals, extracted by the 2015 {\it Planck} likelihood tool\cite{PlanckXI2015}. This is obtained through an hybrid approach based on a direct calculation of the likelihood at large angular scales ($\ell < 30$) and on the use of pseudo-spectral estimates at small angular scales. In Fig. \ref{Spectra} the red curve is the theoretical $\Lambda$CDM model based on the {\it Planck} 2015 best fit of only temperature data and polarization data at large angular scales. The level of agreement with the EE and TE spectra at intermediate and small angular scales is impressive.

\section{Cosmological parameters}

\begin{table}
\centering
\begin{tabular}{|ccccc|}
\hline
\tiny Parameter & \tiny TT+lowP & \tiny TT+lowP+lensing & \tiny TT,TE,EE+lowP & \tiny TT+lowP+lensing+ext \\
\hline
\tiny $\Omega_b h^2$ & \tiny $0.02222 \pm 0.00023$ & \tiny $0.02226 \pm 0.00023$ & \tiny $0.02225 \pm 0.00016$ & \tiny $0.02227 \pm 0.00020$ \\
\tiny $\Omega_c h^2$ & \tiny $0.1197 \pm 0.0022$ & \tiny $0.1186 \pm 0.0020$ & \tiny $0.1198 \pm 0.0015$ & \tiny $0.01184 \pm 0.0012$ \\
\tiny $100\theta_{MC}$ & \tiny $1.04085 \pm 0.00047$ & \tiny $1.04103 \pm 0.00046$ & \tiny $1.04077 \pm 0.00032$ & \tiny $1.04106 \pm 0.00041$ \\
\tiny $\tau$ & \tiny $0.078 \pm 0.019$ & \tiny $0.066 \pm 0.016$ & \tiny $0.079 \pm 0.017$ & \tiny $0.067 \pm 0.013$ \\
\tiny $ln(10^{10}A_s)$ & \tiny $3.089 \pm 0.036$ & \tiny $3.062 \pm 0.029$ & \tiny $3.094 \pm 0.034$ & \tiny $3.064 \pm 0.024$ \\
\tiny $n_s$ & \tiny $0.9655 \pm 0.0062$ & \tiny $0.9677 \pm 0.0060$ & \tiny $0.9645 \pm 0.0049$ & \tiny $0.9681 \pm 0.0044$ \\
\tiny $H_0$ & \tiny $67.31 \pm 0.96$ & \tiny $67.81 \pm 0.92$ & \tiny $67.27 \pm 0.66$ & \tiny $67.90 \pm 0.55$ \\
\tiny $\Omega_m$ & \tiny $0.315 \pm 0.013$ & \tiny $0.308 \pm 0.012$ & \tiny $0.3156 \pm 0.0091$ & \tiny $0.3065 \pm 0.0072$ \\
\tiny $\sigma_8$ & \tiny $0.829 \pm 0.014$ & \tiny $0.8149 \pm 0.0093$ & \tiny $0.831 \pm 0.013$ & \tiny $0.8154 \pm 0.0090$ \\
\tiny $10^9A_se^{-2\tau}$ & \tiny $1.880 \pm 0.014$ & \tiny $1.874 \pm 0.013$ & \tiny $1.882 \pm 0.0012$ & \tiny $1.873 \pm 0.011$ \\
\hline
\end{tabular}
\caption{Parameter 68\% confidence limits for the base $\Lambda$CDM model from {\it Planck} CMB power spectra, in combination with lensing reconstruction ("lensing") and external data ("ext",i.e. BAO+JLA+H0)\cite{PlanckXIII2015}.}
\label{tab:parameters}
\end{table}

The {\it Planck} 2015 likelihood, which is based on the full mission data, is used to explore the cosmological parameter space with a Marcov Chain Monte Carlo (MCMC) method in order to extract the values and uncertainties of the 6 parameters of the $\Lambda$CDM model\cite{PlanckXIII2015}. The results, considering different sets of data, are shown in Table \ref{tab:parameters} including also derived parameters. They represent a great confirmation of the standard cosmological model. They are in excellent agreement with the previous release and, with the addition of external datasets such as measurements of the Baryon Acoustic Oscillations (BAO), type Ia supernovae observations collected in the Joint Light-curve Analysis (JLA) and Hubble constant (H0) measurements, these results represents, to date, the best constraints of all the parameters of the $\Lambda$CDM model.

The same approach can be applied to models that are extensions of the standard cosmological model obtaining for examples tight constraints of the curvature of the Universe ($\Omega_{K}=0.0008^{+0.0040}_{-0.0039}$) confirming the assumption of a flatness, the total sum of the neutrino masses ($\sum m_{\nu} [ev] < 0.194$), the total number of neutrino species ($N_{eff} = 3.04 \pm 0.33$), the ratio between the amplitudes of tensor and scalar perturbations ($r_{0.002} < 0.113$) and the $w$ parameter of Dark energy equation of state ($w=-1.019^{+0.075}_{-0.080}$). 


\section{Lensing}

\begin{figure}[t]
\centering
\includegraphics[scale=0.5]{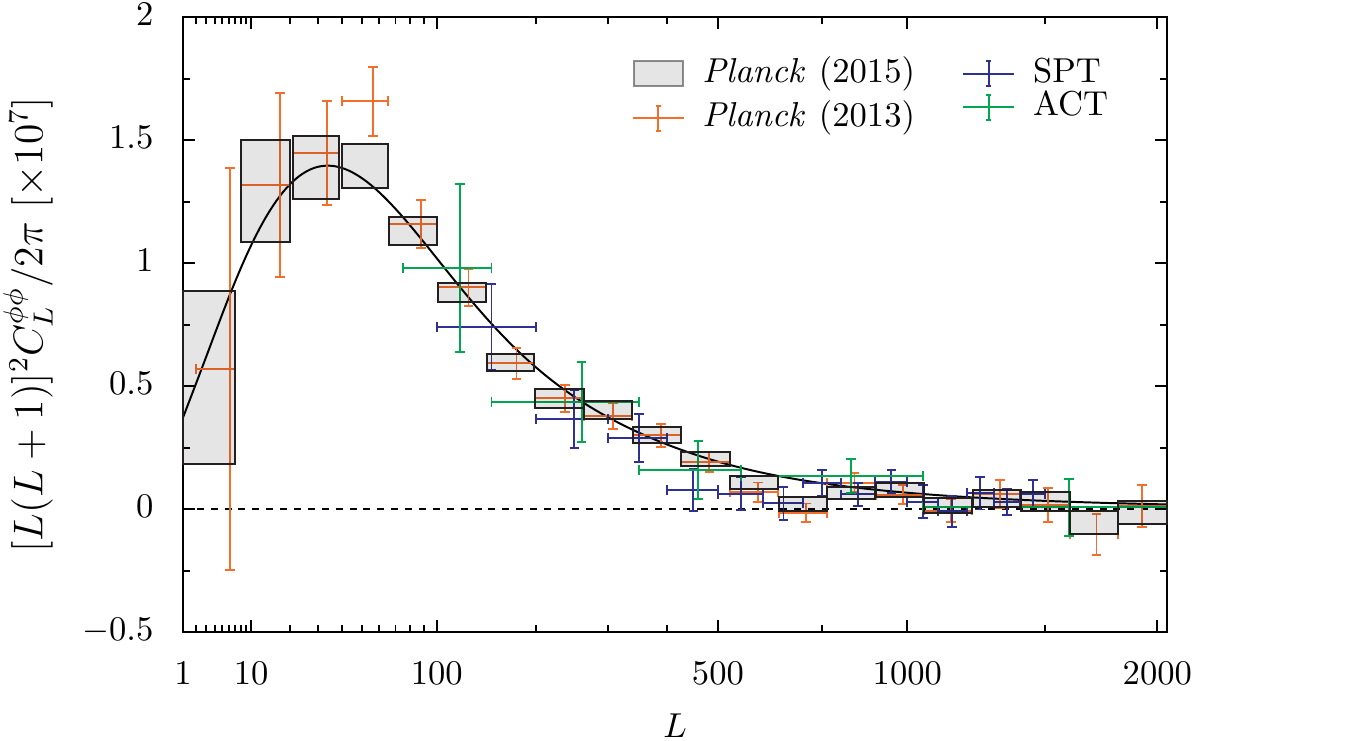}
\caption{{\it Planck} 2015 full-mission lensing potential power spectrum measurement compared with the previous release ({\it Planck} 2013) and with the results of the South Pole telescope (SPT) and the Atacama Cosmology Telescope (ACT)\cite{PlanckXV2015}.}
\label{Lensing}
\end{figure}

During their journey from the last scattering surface to us, the CMB photons experienced small deviations in their path due to the gravitational effect of the matter structures present in the Universe. These deviations cause distinctive statistical signatures onto the observed CMB fluctuations resulting in small statistical anisotropies on the CMB maps. These signatures can be extracted to derive the CMB lensing potential, $C_{\ell}^{\phi\phi}$. Using the SMICA CMB cleaned map and applying five different quadratic estimators based on the correlations of the CMB temperature and polarization, {\it Planck} measured the CMB lensing potential with unprecedented precision \cite{PlanckXV2015}. The broadband amplitude of $C_{\ell}^{\phi\phi}$, shown in Fig. \ref{Lensing}, is now measured to better than 2.5\% accuracy, with a detection of the lensing effect at more than 40$\sigma$. These results allowed to form a full-sky reconstruction of the projected mass distribution. Moreover, lensing B-modes are detected at 10$\sigma$, both through a correlation analysis with the Cosmic Infrared Background (CIB) and via the TTEB 4-point function.

\section{Conclusions and challenges for future generation of instruments}

\begin{figure}[t]
\centering
\includegraphics[scale=0.45]{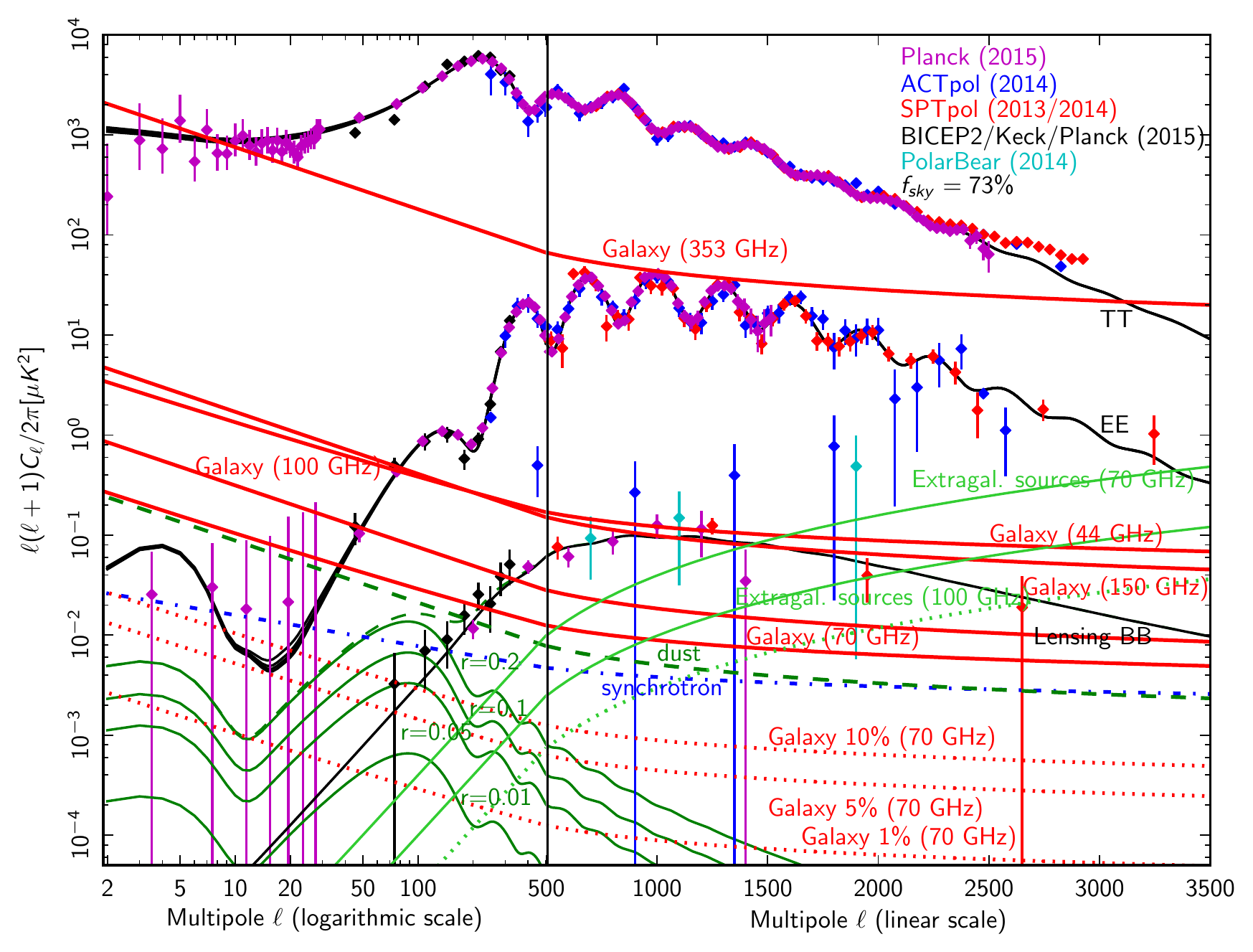}
\caption{To date combined view of the CMB APS measurements in temperature and polarization of the most recent CMB experiments. They are compared with the total foreground emission (synchrotron plus dust) in polarization at different frequencies. The dashed red lines representing various levels of the Galactic Foreground residuals at 70 GHz after an hypothetical component separation are compared to the primordial B modes for various tensor-to-scalar ratios (green lines).}
\label{Forecasts}
\end{figure}

{The {\it Planck} results represent a rich harvest of extremely precise data for both cosmology, providing the best confirmation of the standard cosmological model, and astrophysics, releasing a big reservoir of data that will be analysed in the following years by the scientific community. However, there are still some opened questions. One of the most important is the measurement of the primordial B modes that would be an important confirmation of the inflationary scenario. However, a measurement of the B modes is not only a technological challenge, to produce more stable and sensitive detectors, but also it is an astrophysical and cosmological challenge, because of the presence of foreground emissions that must be removed with great accuracy. In Fig. \ref{Forecasts} we show the measurements of the CMB APS in temperature an polarization made by the most recent CMB experiments. In addition we show the total Galactic foreground contamination in polarization at different frequencies as the sum of synchrotron and dust emission, after a mask that covers the 27\% of the sky has been applied, as described in \cite{PlanckX2015}. The results show that the frequencies around 70 GHz are the cleanest. If future experiments will like to measure the BB primordial spectrum with a tensor-to-scalar ratio, $r$, smaller than 0.01, the component separation will have a great challenge to remove the foregrounds leaving residuals for less than 1\% of the initial amplitude.

\section*{Acknowledgments}

We acknowledge the use of the NASA Legacy Archive for Microwave Background Data Analysis (LAMBDA) 
and of the ESA {\it Planck} Legacy Archive.
We acknowledge partial support by ASI/INAF Agreement 2014-024-R.0 for the {\it Planck} LFI Activity of Phase E2.
Some of the results in this paper have been derived using the HEALPix\cite{Gor05} package.


\end{document}